\documentclass[aps,amsmath,amssymb,preprint,citeautoscript]{revtex4}

\usepackage{amsmath}
\usepackage{graphicx}

\newcommand\ubu{{\overline{u}u}}
\newcommand\dbd{{\overline{d}d}}
\newcommand\ubd{{\overline{u}d}}
\newcommand\dbu{{\overline{d}u}}

\begin{document}

\title{Modeling pion physics in the $\epsilon$-regime of two-flavor QCD \\
using strong coupling lattice QED}
\author{D. J. Cecile and Shailesh Chandrasekharan}
\affiliation{Department of Physics, Box 90305, Duke University,
Durham, North Carolina 27708, USA}

\begin{abstract}

In order to model pions of two-flavor QCD we consider a lattice field theory 
involving two flavors of staggered quarks interacting strongly with $U(1)$ 
gauge fields. For massless quarks, this theory has an 
$SU_L(2)\times SU_R(2) \times U_A(1)$ symmetry. By adding a four-fermion term 
we can break the $U_A(1)$ symmetry and thus incorporate the physics of the QCD
anomaly. We can also tune the pion decay constant $F$, to be small 
compared to the lattice cutoff by starting with an extra fictitious dimension, 
thus allowing us to model low energy pion physics in a setting similar to 
lattice QCD from first principles. However, unlike lattice QCD, a major 
advantage of our model is that we can easily design efficient algorithms 
to compute a variety of quantities in the chiral limit. Here we show that the 
model reproduces the predictions of chiral perturbation theory in the 
$\epsilon$-regime.
\end{abstract}

\maketitle

\section{Introduction}

One of the outstanding problems in lattice QCD is to compute low energy 
hadronic observables, which are dominated by the physics of light quarks, 
with controlled errors. Unfortunately, with current algorithms it is 
difficult to perform calculations at realistic quark masses. In particular, 
one would like to have light pions of mass 140 Mev, which means 
$\frac{m_l}{m_s} \sim \frac{1}{25}$, where $m_l$ refers to the average 
up and down quark masses and $m_s$ refers to the strange quark mass.
In practice the lightest pions with current algorithms can reach about 
250-300 MeV with Wilson type fermions \cite{Giusti:2007hk} and slightly 
lighter with staggered fermions. For example the MILC group has attained 
$\frac{m_l}{m_s} \sim \frac{1}{8}$ with staggered fermions while 
$\frac{m_l}{m_s} > \frac{1}{2}$ is common \cite{Sachrajda:2006wp}.
Given that calculations are usually performed at unphysically large 
quark masses, chiral extrapolations, based on chiral perturbation theory, 
are used to obtain results at realistic quark masses. However, for such 
an approach to be reliable, we must know the range of quark masses over which 
the chiral expansion used in the extrapolation is valid. Unfortunately, this 
is not yet well understood. 

One way to check if chiral extrapolations are reliable is to verify them
in qualitatively distinct regimes. Fortunately, while there are many such 
regimes depending on the values of the pion masses and the physical box sizes, 
the most popular examples are the $p$-regime and the $\epsilon$-regime.
Interestingly, in all the regimes the extrapolation formulas depend on 
the same low energy constants that describe the chiral Lagrangian. 
Hence if the data from a lattice calculation can be fit in both the regimes 
with the same low energy constants one would gain more confidence in the 
extrapolations. As far as we know such comparisons have not yet been 
made in the context of lattice 
QCD. In fact, although chiral perturbation theory is a widely understood and 
accepted tool, much less effort has gone into describing the low energy
physics of a fundamental lattice field theory using the chiral Lagrangian.
Apart from lattice QCD, as far as we know, the only known example where 
lattice results were understood with chiral perturbation theory was in the 
context of the non-linear sigma models \cite{Gockeler:1991sj,Hasenfratz:1990fu,
Hasenfratz:1989ux} and quantum spin systems \cite{Beard:1996wj}. However, 
even in such studies comparison between the $p$-regime and the 
$\epsilon$-regimes were never made. Further, these studies are some what 
outdated given the advances in computing resources. More such studies and 
with simpler models that resemble QCD closely may teach us more about chiral 
extrapolations. Motivated by this, we introduce and study a 
lattice field theory model of pions in two-flavor QCD.

Our model is nothing but strongly coupled lattice QED with two flavors of 
staggered fermions. This model exhibits the same symmetries as 
two flavor QCD and was recently used to study the chiral phase transition 
\cite{Chandrasekharan:2006zq}. The strong coupling limit was studied 
considerably in the eighties 
\cite{Blairon:1980pk,Kawamoto:1981hw,Kluberg-Stern:1982bs,
Martin:1982tb,Dagotto:1986xt,Dagotto:1986ms,Dagotto:1986gw,Karsch:1988zx,
Klatke:1989xy,Boyd:1991fb} 
because some of the qualitative physics remains despite this limit
suffering from the worst lattice artifacts.  In fact, even a $U(1)$ gauge 
theory exhibits confinement and chiral symmetry breaking in the strong 
coupling limit. Notice further that the taste symmetry of continuum 
staggered fermions is irrelevant since it is maximally broken at strong 
couplings. The relevant symmetry is due to the two flavors and resembles 
QCD closely. Thus, our model allows us to model pion physics from a 
fundamental lattice field theory, very similar to QCD.

Since the pion is the lightest QCD bound state and plays a crucial role in 
chiral symmetry breaking, even a qualitative description of it from first
principles may prove useful.
However, unless we find a way to fine tune our model, the pion decay 
constant $F$ (defined here in the chiral limit), is naturally close to the 
cutoff. Thus, the low energy pion physics in our model may be described 
by a chiral Lagrangian that contains information about lattice artifacts. 
In order to circumvent this 
problem, we define our model in $d+1$ dimensions where $d=4$ is the space 
time dimensions. The extra dimension plays the role of a fictitious 
temperature which allows us to tune $F$ to values much smaller than 
the cut-off. Thus, we can still explore ``continuum-like'' physics even 
in the strong coupling limit.

In addition to its similarity with QCD, another important motivation for 
studying the above model is that the gauge dependent degrees of freedom 
can be integrated over, which significantly simplifies the theory. Further, 
a new class of algorithms, called {\em the Directed Path Algorithm}, allows 
us to study the chiral limit very efficiently \cite{Adams:2003cc}. Here we 
extend the algorithm to our model and show that the results in the chiral 
limit agree with predictions in the $\epsilon$-regime. A preliminary version 
of this work can be found in \cite{Cecile:2006zr}.

\section{Model and Symmetries}

The Euclidean space action of the model we consider is given by (note that 
the usual factor of $\frac{1}{2}$ in the fermion kinetic term has
been absorbed into the fields):
\begin{equation}
S = - \sum_{x}\sum_{\mu=1}^{d+1}
\eta_{\mu,x}\bigg[\mathrm{e}^{i\phi_{\mu,x}}{\overline\psi}_x
{\psi}_{x+\hat\mu}
-\mathrm{e}^{-i\phi_{\mu,x}}{\overline\psi}_{x+\hat\mu}{\psi}_x\bigg]
-\sum_x \bigg[m{\overline\psi}_x{\psi}_x
+\frac{\tilde c}{2}\bigg({\overline\psi}_x{\psi}_x \bigg)^2\bigg],
\label{eq1}
\end{equation}
where $x$ denotes a lattice site on a $d+1$ dimensional hyper-cubic lattice
$L_t \times L^d$.  Here $L_t$ represents a fictitious time direction and
will be used to tune $F$, the non-perturbative physical mass scale, to 
be small compared the lattice cutoff. On the other hand $L^d$ will be the
usual Euclidean space-time box. $\overline\psi_x$ and $\psi_x$ are two 
component Grassmann fields that represent the two quark $(u,d)$ flavors of 
mass $m$, and $\phi_{\mu,x}$ is the compact $U(1)$ gauge field through 
which the quarks interact. Here $\mu = 1,2,...,d,d+1$  runs over the $d+1$ 
directions. The $\mu=1$ direction will denote the fictitious temperature 
direction, while the remaining directions represent Euclidean space-time.
The usual staggered fermion phase factors $\eta_{\mu,x}$ obey the relations: 
$\eta_{1,x}^2 = T$ and $\eta_{i,x}^2 = 1$ for $i=2,3,...,d+1$. The parameter 
$T$ controls the fictitious temperature and will be used to tune 
to the continuum limit ($F \ll 1$). The coupling $\tilde c$ will set 
the strength of the anomaly.

When $\tilde c,m = 0$, the action exhibits a global 
$SU_L(2)\times SU_R(2)\times U_A(1)$ symmetry like $N_f=2$ QCD. To see this 
it is first useful to note that every lattice site can be classified as 
an {\it even} or {\it odd} site. Then it is easy to see that the action 
is invariant under the following transformations: ${\overline{\psi}_o} 
\rightarrow {\overline{\psi}_o}V_{L}^{\dagger} \exp(i\phi), {\psi_o} 
\rightarrow  \exp(i\phi)V_{R}{\psi_o}, {\overline{\psi}_e} 
\rightarrow {\overline{\psi}_e}V_{R}^{\dagger}\exp(-i\phi), {\psi_e} 
\rightarrow \exp(-i\phi)V_{L}{\psi_e}$.
Here $V_L$ and $V_R$ are $SU(2)$ matrices and can be parametrized by:
$\exp(i\vec{\theta}\cdot\vec{\sigma})$ where $\sigma_i$ is a Pauli matrix
that acts on the flavor space. At $\tilde c\neq 0$, $U_A(1)$ is explicitly 
broken and the action is invariant under $SU_L(2) \times SU_R(2) \times Z_2$. 
Thus, the coupling $\tilde c$ induces the effects of the anomaly. Further 
at $m\neq 0$, it is necessary to set $V_L=V_R$ for the action to
remain invariant. Thus, with a mass term the chiral symmetry 
$SU_L(2)\times SU_R(2)$ is explicitly broken down to $SU_V(2)$. In order
to mimic the real world with $u$,$d$ quarks we need to set 
$\tilde c\neq0$ and $m\neq0$. Thus, our model 
has the same chiral symmetry as $N_f=2$ QCD. Further, based on previous 
mean field strong coupling calculations \cite{Kluberg-Stern:1982bs}, 
we expect that the symmetry breaking pattern is also similar to full QCD. 

\section{Monomer-Dimer-Pion Loop-Instanton Representation}
\begin{figure}[htb]
\begin{center}
\includegraphics[width=12cm]{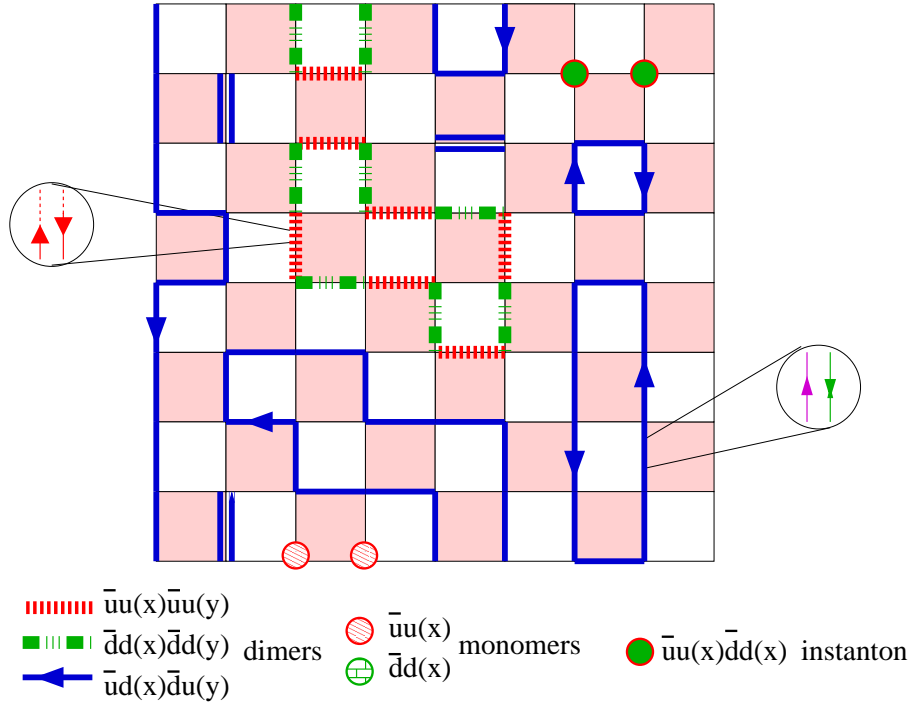}
\end{center}
\caption{\label{fig0}An example of a $2\times 2$ lattice configuration
as discussed in the text. }
\end{figure}

The partition function of our model is equivalent to that of 
a classical statistical mechanics model involving configurations made up of 
gauge invariant objects such as monomers, dimers, pion loops and 
instantons \cite{Rossi:1984cv}. We call these MDPI configurations and 
denote the set of all such configurations by ${\cal K}$. Each MDPI 
configuration is characterized by three site variables $I(x),n_u(x),n_d(x)$
and three bond variables $\pi^u_\mu(x),\pi^d_\mu(x),\pi^1_\mu(x)$.
Here $n_u(x)$ is the number of $\ubu$ monomers, $n_d(x)$ the number of $\dbd$ 
monomers and $I(x)$ the number of instantons (or $\ubu\dbd$ double-monomers) 
associated with the site $x$. On the other hand $\pi_{\mu}^u(x)$ denotes the 
number of $\ubu\ubu$ dimers, $\pi_{\mu}^d(x)$ the number of $\dbd\dbd$ dimers, 
and $\pi_{\mu}^1(x)$ the number of oriented ($\ubd\dbu$ or $\dbu\ubd$) dimers 
that live on the bond connecting $x$ and $x+\hat{u}$. In our notation 
$\pi^u_{-\mu}(x) = \pi^u_\mu(x-\hat{\mu})$ and similarly for other bond
variables. The allowed values for these variables are: 
\begin{equation}
I(x) = 0,1 \quad n_d(x) = 0,1 \quad n_u(x) = 0,1 \quad
\pi_{\mu}^d(x) = 0,1 \quad \pi_{\mu}^u(x) = 0,1 \quad 
\pi_{\mu}^1(x) = -1,0,1
\label{eq8}
\end{equation}
Note that $n_u(x)=n_d(x)=1$ is not allowed since it is absorbed into
$I(x)$. Due to the Grassmann nature of the fermion fields the following 
constraints must also be satisfied at each site $x$:
\begin{subequations}
\begin{eqnarray}
\sum_{\mu} \pi_{\mu}^1(x) &=& 0 \\
2I(x) + \sum_{\mu}\bigg[\pi_{\mu}^u(x) + \pi_{\mu}^d(x)+ n^u(x) 
+ n^d(x)\bigg] + \sum_{\mu}\big|\pi_{\mu}^1(x)\big| &=& 2 \\
n_u(x) + \sum_{\mu}\bigg[\pi_{\mu}^u(x) - \pi_{\mu}^d(x)\bigg] 
- n_d(x) &=& 0
\end{eqnarray}
\end{subequations}
where the sum over $\mu$ goes over $\pm1,...\pm(d+1)$. Fig.(\ref{fig0}) 
gives an illustration of an MDPI configuration in $1+1$ dimensions. 

Note that $I(x)=1$ on a site $x$ breaks the $U_A(1)$ symmetry but not the 
$SU_L(2)\times SU_R(2)$ symmetry and hence it is called an {\em instanton}. 
In addition to these, each configuration can contain {\em neutral pion-loops} 
of alternating $\ubu\ubu$ and $\dbd\dbd$ dimers or {\em charged pion-loops} 
of $\ubd\dbu$ (or $\dbu\ubd)$ dimers. Neutral pions can also
form open strings at non-zero $m$ with $\ubu$ or $\dbd$ monomers at 
the ends, while charged pions cannot form such open strings in our model.
Note also that if a bond consists of two dimers (double bond) it can be 
considered 
both as a neutral pion-loop or a charged pion-loop. Finally, it is useful 
to recognize that charged pion-loops are naturally oriented while neutral 
pion-loops are not. On the other hand it is possible to give an orientation 
to the neutral pion-loops also. Here we will assume that a $\ubu\ubu$ dimer 
on a forward bond from an even site and a $\dbd\dbd$ dimer on a forward bond 
from an odd site are considered oriented forward. Similarly a $\ubu\ubu$ dimer 
on a forward bond from an odd site and a $\dbd\dbd$ dimer on a forward bond 
from an even site are considered oriented backward. With this convention it 
is clear that neutral pion-loops are also oriented. This will be useful when
constructing the algorithm. If we define $c\equiv \tilde{c}+m^2$, the 
partition function is given by:
\begin{equation}
Z = \sum_{[{\cal K}]} \prod_{x} m^{n_d(x)} m^{n_u(x)}c^{I(x)}.
\label{pf}
\end{equation}
where the sum includes all allowed MDPI configurations.

\section{Algorithm}

Since the partition function in Eq.~(\ref{pf}) is a sum over positive 
definite weights, a Monte-Carlo algorithm can be used to sample the 
MDPI configurations ${\cal K}$. We have extended the work of 
\cite{Adams:2003cc,Chandrasekharan:2006tz} and constructed the 
algorithm. It is comprised of four updates : 
a {\em $u \leftrightarrow d$ flip}, a {\em loop swap}, and 
a {\em directed-path mass update} and a {\em directed-path fixed-monomer 
update}. While each update by itself satisfies detailed balance, one needs 
a combination of these updates to ensure ergodicity. The first
three updates are sufficient to ensure ergodicity in general.
The fixed monomer update in combination with the flip and the swap
allows us to calculate quantities in a fixed monomer sector which will 
be useful in some of our results. We now discuss each update in detail.  

{\em The $u \leftrightarrow d$ flip update}
This update changes a $u$ quark to a $d$ quark and vice-versa on a
pion loop or string. Thus, a $\ubu\ubu$ dimer becomes a $\dbd\dbd$
dimer and vice-versa or a $\ubd\dbu$ dimer becomes a $\dbu\ubd$ dimer
and vice-versa. Similarly a $\ubu$ monomer becomes a $\dbd$ monomer.
The complete update is as follows:
\begin{enumerate}
\item Let $V$ denote the set of all lattice sites. A lattice site $x_0\in V$ 
is selected randomly.
\item If $x_0$ is part of a pion loop or a string, the $u\leftrightarrow d$
flip is performed on all dimers and monomers associated with the loop or 
string. The update then ends.
\item If $x_0$ is part of an instanton or a double dimer, the update ends.
\end{enumerate}
Because the configuration before and after the update have equal weights, 
this update automatically satisfies detailed balance.

{\em Loop swap update.}
This update swaps a neutral pion-loop into a charged pion-loop and vice-versa. 
The complete update is as follows:
\begin{enumerate}
\item A lattice site $x_0\in V$ is selected randomly.
\item If $x_0$ is part of a neutral pion-loop, the complete loop is 
changed into a charged pion-loop and vice-versa.  The update then ends.
\item If $x_0$ is not part of a loop, the update ends.
\end{enumerate} 
Again, because neutral and charged pion-loops have equal weights, this update 
automatically satisfies detailed balance.

{\em Directed-path mass update.}
This update is more complex and can create and destroy monomers and 
instantons while at the same time it can change the shape of the pion 
loops. We construct and use two types of directed-path update that differ 
on the sites they are allowed to touch. The {\em charged-pion
directed path update} can only touch sites containing either charged 
pion-loops (including double dimers) and instantons, while the
{\em neutral-pion directed path update} can only touch sites containing 
neutral pion-loops (including double dimers), instantons and monomers. 
Since the rules of the update are almost the same we will discuss
them together. We will call the set of sites a particular update is allowed 
to touch as ${\cal C}$. 

To understand the directed path update it is useful to think of the MDPI
configurations as constructed with oriented objects. As we have already 
discussed above every loop is oriented. Open strings of neutral pions are 
also oriented with an incoming monomer and an outgoing monomer at the two 
ends. An instanton is also an oriented string of zero length with a forward 
and a backward monomer on the same site. A forward monomer is oriented into 
the $\mu=0$ direction and a backward monomer is oriented 
from the $\mu=0$ direction. Thus, each MDPI configuration naturally gives 
every site a forward and a backward direction. The directed-path update 
essentially updates these directions on the sites it touches.

\begin{figure}[htb]
\vskip0.5in
\begin{center}
\includegraphics[width=12cm]{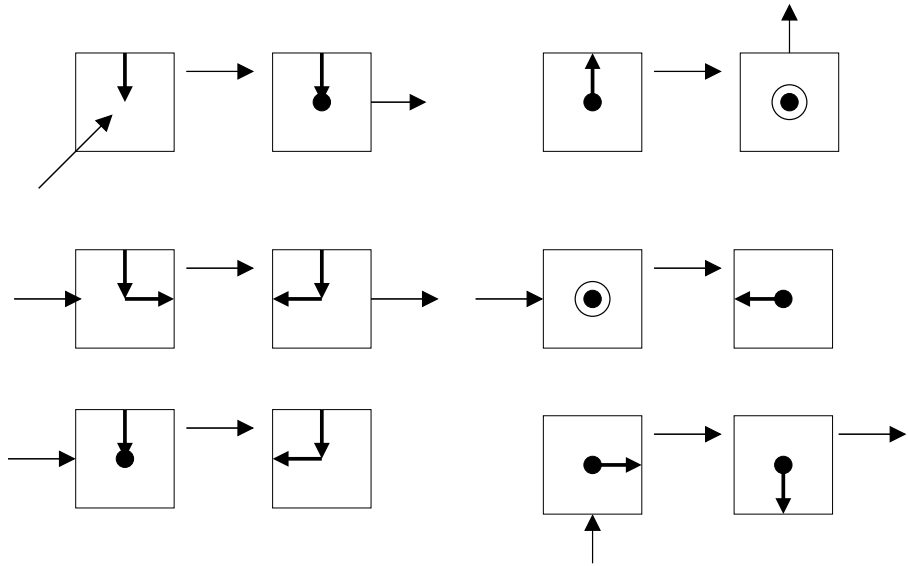}
\end{center}
\caption{\label{activ} Examples of the active update at a site. The 
absence of an outgoing arrow implies that the update remains on the site.}
\end{figure}

\begin{figure}[htb]
\vskip0.5in
\begin{center}
\includegraphics[width=12cm]{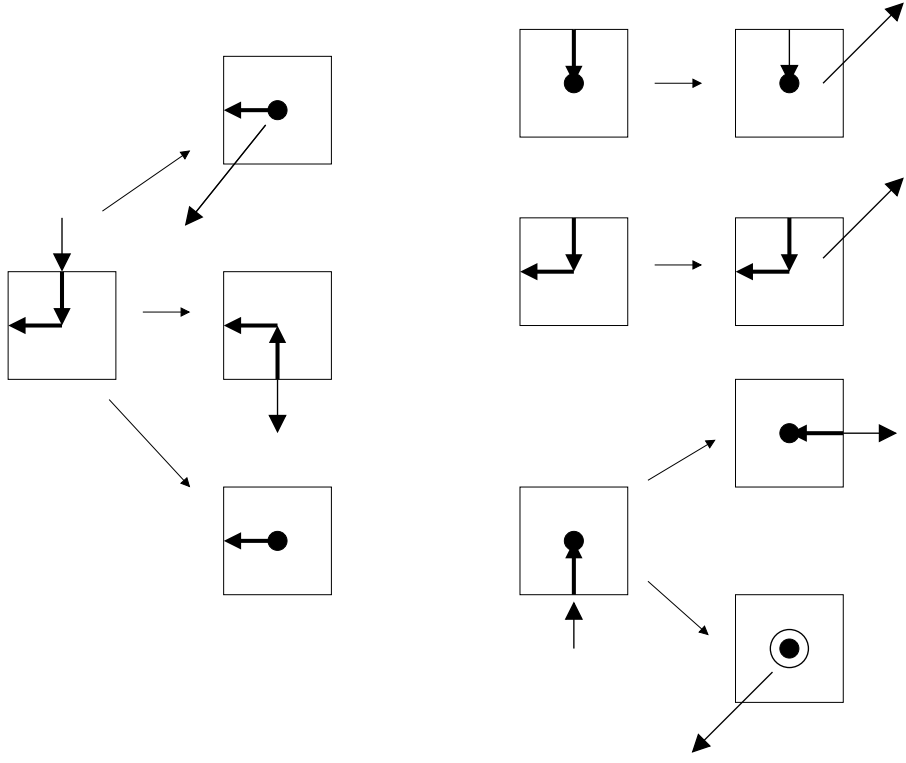}
\end{center}
\caption{\label{passiv} Examples of the passive update at a site. The 
absence of an ingoing or an outgoing arrow implies that the update is 
on the site.}
\end{figure}

The update consists of two parts: an {\em active update} and a {\em 
passive update}. During an {\em active update} one replaces the forward 
direction $\mu$ at the site $x$ by the direction $\nu$ through which the
site was approached. If $\mu \neq 0$, then after replacing $\mu$ with $\nu$, 
the update moves to the neighboring site $x+\hat{\mu}$ and replaces the 
backward direction on the new site with zero. Otherwise the update remains 
on the same site $x$ without disturbing the backward direction on it.
Some examples of the active update is shown in Fig.(\ref{activ}).
During a {\em passive update} if the backward direction $\mu \neq 0$,  
the update ends. Otherwise a new backward direction $\nu$ is chosen with 
probability $w_{\nu,x}/W$. Here $w_{\nu,x}=0$ if $x+\hat{\nu} \notin {\cal C}$,
otherwise $w_{\nu=0,x}=c'$, $w_{\nu=\pm 1}=T$ and $w_{\nu,x}=1$ otherwise. 
$W$ is the sum of weights of all possible choices. One of the $\nu$ is chosen 
according to heat bath probabilities.  When $\nu=0$ is chosen then either
a monomer or an instanton can be created. If the forward direction on the 
site is non-zero then the probability of creating a monomer is $m^2/c'$ 
and creating an instanton is $c/c'$ where $c'=c+m^2$. If the forward direction
is zero then only an instanton can be created and so $c'=c$.
 The update ends on a passive site if $\nu=0$ is chosen with a 
creation of a monomer. On the other hand if an instanton is created the 
update can end only if the forward direction on the site was also zero.
Otherwise the update remains on that site and continues. If $\nu \neq 0$ 
is chosen then the backward direction is changed to $\nu$ and the 
update moves to the neighboring site $x+\hat{\nu}$. Some examples of the
passive update is shown in figure \ref{passiv}. Note that in the charged 
pion directed loop update $m^2$ is assumed to be zero so monomers can not be 
generated.

The complete update is simple and is described below.
\begin{enumerate}
\item A lattice site $x_0\in V$ is selected randomly. If $x_0$ does not belong 
to ${\cal C}$ the whole update ends. Further, if both the forward and the 
backward directions at this site are non-zero, the site is active and was 
reached from the $\nu = 0$ direction.  Otherwise the site is assumed to be 
passive.
\item Each current site is updated using the rules of an active site or a
passive site successively until the update ends.
\end{enumerate}

{\em Directed path fixed monomer update.}
This update is identical to the {\em directed path mass update}, except for 
a minor change in the {\em neutral pion directed path update}. The change is 
such that $\ubu$ or $\dbd$ monomers are not allowed to be created or 
annihilated during the passive update. Thus, the update allows the 
monomers to change positions while keeping the total monomer fixed. Note that 
instanton number can of course change. 

\section{Observables}
\label{obs}
Numerous observables can be measured with our algorithm. The simplest are
the three helicity moduli or current susceptibilities.  In particular, for 
a conserved current $J_{\mu}^i(x)$, the helicity modulus 
(current susceptibility) is defined as:  
\begin{equation}
Y_i = \frac{1}{d L^d}\bigg\langle 
\sum_{\mu=1}^d\bigg(\sum_x J_{\mu}^i(x)\bigg)^2\bigg\rangle
\end{equation}
where we are assuming a $L_t \times L^d$ lattice. There are three 
conserved currents in our model. They are the axial, chiral, and vector 
currents which are given by:
\begin{subequations}
\begin{eqnarray}
J_{\mu}^a(x) &=& (-1)^x\big[\pi_{\mu}^u(x)+\pi_{\mu}^d(x)+|\pi_{\mu}^1(x)|
\big] \\
J_{\mu}^c(x) &=& (-1)^x\big[\pi_{\mu}^u(x)-\pi_{\mu}^d(x)\big] \\
J_{\mu}^v(x) &=& \pi_{\mu}^1(x) \\\nonumber
\end{eqnarray}
\end{subequations}
The current susceptibilities $Y_a,Y_c$ and $Y_v$ are diagonal observables 
in our approach and can be easily computed for every configuration that 
contributes to the partition function. 

We can also measure two point correlation functions of many fermion bilinears.
In particular we can compute
\begin{subequations}
\begin{eqnarray}
G^a_\pi(x,y) &=& \frac{1}{2}
\langle \overline{\psi}_x i\sigma^a (-1)^x\psi_x \ 
\overline{\psi}_y i\sigma^a(-1)^y\psi_y\rangle 
\\
G_\sigma(x,y) &=& \frac{1}{2}
\langle \overline{\psi}_x\psi_x \ \overline{\psi}_y\psi_y\rangle 
\\
G_\eta(x,y) &=& \frac{1}{2}
\langle \overline{\psi}_x i(-1)^x\psi_x\  
\overline{\psi}_y i (-1)^y\psi_y\rangle
\\
G^a_\delta(x,y) &=& \frac{1}{2}
\langle \overline{\psi}_x\sigma^a\psi_x \ 
\overline{\psi}_y\sigma^a\psi_y\rangle
\end{eqnarray}
\end{subequations}
These correlation functions are non-diagonal observables and cannot
be measured easily on the configurations that contribute to the path
integral. However, they can be measured during the {\em charged-pion 
directed path update} as follows:
\begin{enumerate}
\item All the $G_i(x,y)$ are set to zero before the update.
\item If $x$ is the first active site and $y$ is one of the passive 
sites visited during the directed path update, then $G_i(x,y)$ is changed 
as follows:
\begin{subequations}
\begin{eqnarray}
G_\pi(x,y) &=& G_\pi(x,y) + \frac{L_t L^d}{2 W_y} \delta_{x,x_0}\delta_{y,y_0} \\
G_\delta(x,y) &=& G_\delta(x,y) - (-1)^{x+y}\frac{L_t L^d}{2 W_y} \delta_{x,x_0}\delta_{y,y_0}\\
G_\sigma(x,y) &=& G_\pi(x,y) + \frac{L_t L^d}{2 W_y} \delta_{x,x_0}\delta_{y,y_0}\\
G_\eta(x,y) &=& G_\delta(x,y) - (-1)^{x+y}\frac{L_t L^d}{2 W_y} \delta_{x,x_0}\delta_{y,y_0}
\end{eqnarray}
\end{subequations}
If $n_m(z)$ is the number of monomers and $n_I(z)$ is the number
of instantons at any lattice site $z$ then the $G_\eta$ and $G_\sigma$
correlator get additional ``disconnected'' contributions:
\begin{subequations}
\begin{eqnarray}
G_\sigma(x,z) &=& G_\sigma(x,z) + 
[1 + n_m(z)+ 2 n_I(z) m^2/\tilde{c}]\frac{L_t L^d}{2 W_y}   \\
G_\eta(x,z) &=& G_\eta(x,z) -
(-1)^{x+z} [n_m(z) + 2 n_I(z) m^2/\tilde{c})]\frac{L_t L^d}{2 W_y} \\
\end{eqnarray}
\end{subequations}
\item The function $G_i(x,y)$ thus obtained for each directed path update, 
when averaged over directed path updates, yields the final correlation 
function defined above.
\end{enumerate}

Once the correlation functions are known, the corresponding susceptibilities, 
$\chi_\pi$, $\chi_\sigma$, $\chi_\eta$ and $\chi_\delta$ are given by:
\begin{equation}
\chi_i = \frac{2}{L^d} \sum_{x,y} G_i(x,y)
\end{equation}
for $i=\sigma,\pi,\eta,\delta$.  We have normalized $\chi_i$ such that
\begin{equation}
\label{chiralcond}
\chi_i = \frac{1}{L^d}\frac{1}{Z}\frac{\partial^2 Z}{(\partial m)^2}
\end{equation}
We have tested our algorithm by comparing the results for all the above
observables with exact analytic calculations for each of these observables 
on a $2 \times 2$ lattice. The exact results and those from the algorithm
are shown in Appendix I.

\section{Results in the $\epsilon$-regime}

In order to establish that we can indeed use our approach to model
pions of two flavor QCD, in this paper we focus on the $\epsilon$ regime 
of chiral perturbation theory 
\cite{Neuberger:1987zz,Neuberger:1987fd,Gasser:1987ah,
Gasser:1987zq,Hasenfratz:1989pk,Hansen:1990yg,Hansen:1990un}. 
In the phase where chiral symmetry is broken and
the anomaly is large, the low energy physics of our model must be describable 
by the Euclidean chiral Lagrangian
\begin{equation}
{\cal L} = 
\frac{F^2}{4}
\mathrm{tr}\Big(\partial_\mu U^\dagger \partial_\mu U\Big)
- m \Sigma tr \Big(U + U^\dagger \Big)
\end{equation}
where $F$ is the pion decay constant in the chiral limit and $\Sigma$
is the chiral condensate and $U\in SU(2)$ is the pion field. The $\epsilon$ 
regime involves the limit where $L$, the linear size of the four dimensional 
hypercube, is taken to be large such that $F L \ll 1$ but $m \Sigma L^4$ 
is held fixed. In this limit a variety of quantities have been computed in
the literature. For example the behavior of $\chi_\sigma$ as a function of
$L$ at $m=0$ was obtained in \cite{Hasenfratz:1989pk} for the $O(N)$ model. 
The $N=4$ result, relevant here, is 
\begin{equation}
\label{chipred}
\chi_{\sigma} = \frac{\Sigma^2 L^4}{4} \Bigg[1 + \frac{3 \beta_1}{(FL)^2} +
\frac{a}{(FL)^4}  + ...\Bigg]
\end{equation}
The dependence of $Y_c$ and $Y_v$ were obtained in \cite{Hansen:1990yg} and 
it was shown that
\begin{subequations}
\label{ypred}
\begin{eqnarray}
Y_c &=& \frac{F^2}{2} \Bigg(\bigg\{1 + 
\frac{\beta_1}{(FL)^2} +  \frac{a'}{(FL)^4} + ...\bigg\} + 
\frac{u^2}{24}\bigg\{1 + \frac{3 \beta_1}{(FL)^2} + \frac{b_c}{(FL)^4} 
+ ...\bigg\} + {\cal O}(u^4)\Bigg) 
\nonumber \\ \\
Y_v &=& \frac{F^2}{2} \Bigg(\bigg\{1 + 
\frac{\beta_1}{(FL)^2} +  \frac{a'}{(FL)^4} + ...\bigg\} - 
\frac{u^2}{24}\bigg\{1 + \frac{3 \beta_1}{(FL)^2} + \frac{b_v}{(FL)^4} 
+ ...\bigg\} + {\cal O}(u^4)\Bigg)
\nonumber \\
\end{eqnarray}
\end{subequations}
for small $u = \Sigma m L^4[1 + 3\beta_1/(2 (FL)^2)]$. In the above formulas 
$\beta_1 = 0.14046$ is the shape coefficient and $a$,$a'$ $b_c$,$b_v$ are 
constants that depend on higher order low energy constants of the chiral 
expansion. Also note that $Y_c=Y_v$ when $u=0$ reflects the chiral symmetry 
of the theory.

\begin{figure}[htb]
\vskip0.5in
\begin{center}
\includegraphics[width=12cm]{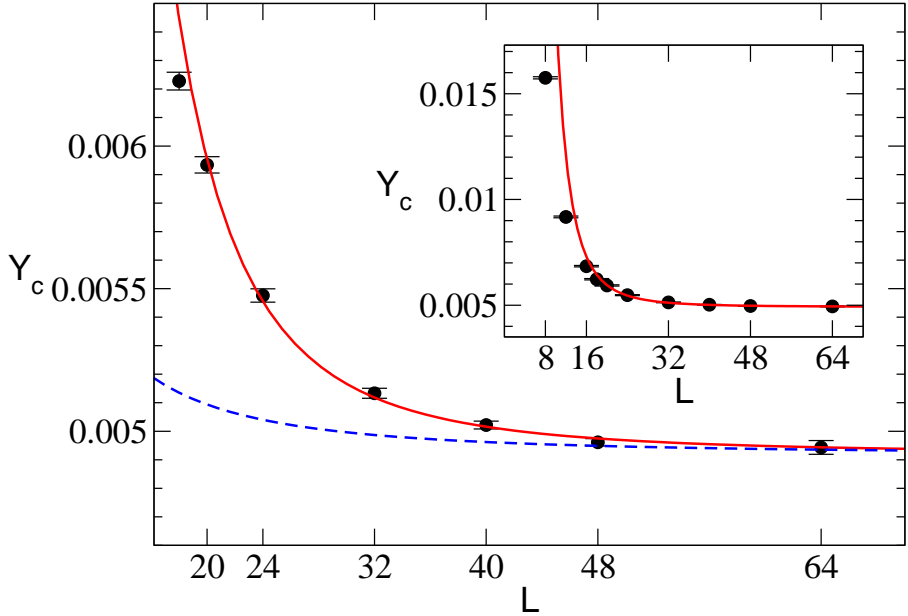}
\end{center}
\caption{\label{fig2} Vector current susceptibility $Y_C$ 
as a function of $L$ at $T=1.733$, $c=0.3$ and $m=0$ [$Y_C = Y_V$].
The solid line shows the fit with $F=0.0992$ and $a'=2.7$.  The dotted
line shows the same curve but with $a'=0$. }
\end{figure}

We now argue that the calculations in our model are consistent with 
Eqs.(\ref{chipred}) and (\ref{ypred}). We choose
$c=0.3$ and $T=1.733$ with fixed $L_t=2$. These parameters are chosen
so that chiral symmetry is spontaneously broken and the anomalous pion 
mass ($M_\eta$) is about $1$ in lattice units. As we will see, this choice 
of $T$ makes $F \sim 0.1$ in lattice units, which should make our results less 
sensitive to lattice artifacts. In Fig.(\ref{fig2}) we plot our 
data for $Y_c$ as a function of $L$ for $m=0$. The solid lines are fits 
to Eq.(\ref{ypred}). The fits are extremely good if we allow for $a'\neq 0$
and use lattice sizes above $L \sim 24$. We can then extract $F = 0.0992(1)$ 
and $a'=2.7(1)$ with a $\chi^2/DOF = 0.8$. However, as can be seen from the 
figure, this means that setting $a'=0$, i.e.. using only the leading 
correction in the chiral expansion, will not be a good approximation for 
$L < 48$. This is clearly due to the smallness of $a'/\beta_1$. In other 
words, although our data is consistent with the Eq.(\ref{ypred}), 
unfortunately we are not yet sensitive to the universal corrections at
order ${\cal O}(1/L^2)$.

\begin{figure}[htb]
\vskip0.5in
\begin{center}
\includegraphics[width=12cm]{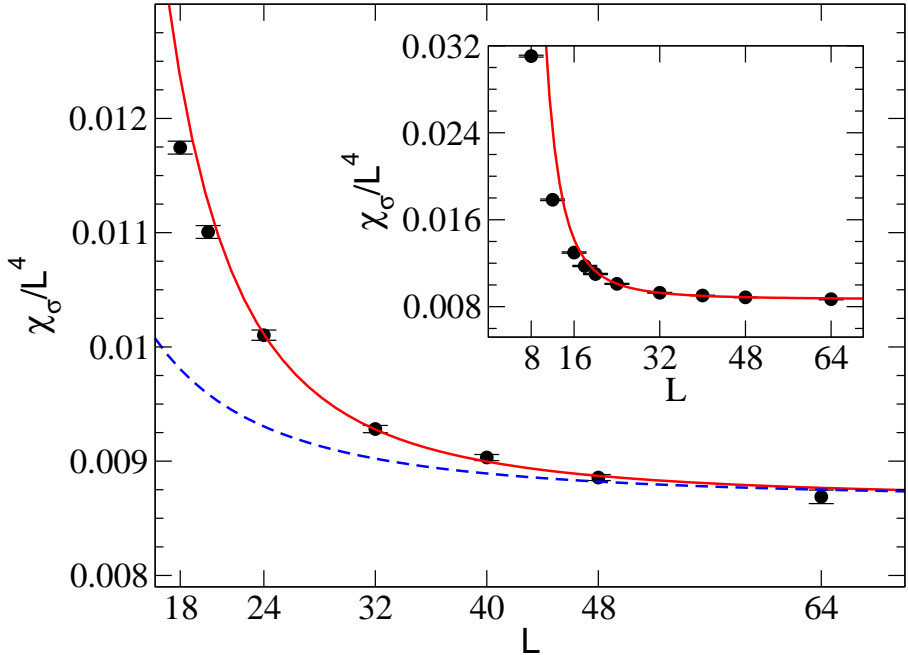}
\end{center}
\caption{\label{fig3} Chiral condensate susceptibility $\chi_\sigma$ 
as a function of lattice size $L$ at $T=1.733$, $c=0.3$ 
and $m=0$ [$\chi_\sigma = \chi_\pi$]. The solid line is the plot of 
Eq.(\ref{chipred}) with $\Sigma = 0.1866$, $F=0.0992$ and $a=3.0$. 
The dotted line shows the same curve but with $a=0$.}
\end{figure}

Let us next focus on the condensate susceptibility $\chi_\sigma$. In 
Fig(\ref{fig3}) we plot $\chi_\sigma/L^4$ as a function of $L$. The solid 
line is a fit to the data using the Eq.(\ref{chipred}) where we fix 
$F=0.0992$ and take only the data for $L > 20$ in the fit. We find 
$\Sigma = 0.1862(2)$, $a = 3.0(2)$ with a $\chi^2/DOF = 1.3$. As can be 
seen again, the universal correction at order ${\cal O}(1/L^2)$ is small 
compared to the next order non-universal correction for $L < 48$.

In the current work we neglect the $\log(L)$ corrections that arise at the 
order $1/(FL)^4$ \cite{Gockeler:1991sj}. The reason is as follows: Consider
for example the chiral condensate susceptibility defined 
Eq.~(\ref{chiralcond}). Using the results of 
\cite{Hasenfratz:1989pk,Gockeler:1991sj,Hasenfratz:1990fu}, 
we can obtain the logarithmic 
corrections to Eq.~(\ref{chipred}). In particular we find that
\begin{equation}
\chi_\pi = \frac{\Sigma^2 L^4}{4} \Bigg[1 + \frac{3\beta_1}{F^2L^2} 
+ \frac{1}{F^4 L^4}
\Big\{\alpha +  \frac{15}{16\pi^2}(\log FL)\Big\}
+ O\big(\frac{1}{F^5L^5}\big) \Bigg] 
\end{equation}
where now $\alpha = (3\beta_1^2 + 15\beta_2)/2 + 
3[\log(\Lambda_M/F) + 4\log(\Lambda_\Sigma/F)]/16\pi^2$ and 
$\beta_2 = -0.020305$ 
is another shape coefficient. The mass scales $\Lambda_M,\Lambda_\Sigma$ 
encode the non-universal information of our model and are defined in 
\cite{Hasenfratz:1989pk}. Assuming $L_1=20$ is the 
smallest lattice size and $L_2= 64$ is the largest lattice size we
use in the fits, we see that the change in the logarithmic correction 
term $15\log(L_2/L_1)/(16\pi^2) \sim 0.1$ is within errors of the 
constant $a=3.0(2)$ obtained above by fitting the $\chi_\pi$ data to 
Eq.(\ref{chipred}). Thus, we believe our errors are still large and we 
are not yet sensitive to the logarithmic corrections. Interestingly,
since $15 \log(FL)/(16\pi^2)$ is much smaller than $a$
in the region of our fits, we can estimate that  
$a \sim 3[\log(\Lambda_M/F)+4\log(\Lambda_\Sigma/F)]/(16\pi^2)$, which 
means that $[\log(\Lambda_M/F)+4\log(\Lambda_\Sigma/F)]$ $(\sim 150)$ 
is unnaturally large, and the factor $\frac{1}{16\pi^2}$ is essential 
to keep the coefficient of $1/(FL)^4$ of order $1$. In other words
factors like $\frac{1}{16\pi^2}$ cannot always be assumed to be small since 
they can be multiplied by large numbers.

Having confirmed that our results are consistent with Eq.(\ref{ypred}) 
when $u=0$, we can also check that our model gives results consistent
with Eq.(\ref{ypred}) at order $u^2$. One way to do this is to tune the 
quark mass and the volume such that $u$ is fixed and small. This is 
cumbersome. For example even at $u=1$ we see that for $L=48$ our quark
mass should be as small as $10^{-6}$. Since $m^2$ is involved in a
probability, one begins to worry about double precision arithmetic.
Thus, here we devise another method. To understand our approach 
let us expand the partition function in powers of the quark mass
\begin{equation}
Z = Z_0 + m^2 Z_2 + m^4 Z_4 + ....
\end{equation}
where $Z_n$ is obtained from configurations with $n$ monomers. In this 
expansion we can neglect the $m^2$ contribution to instanton weights as 
they will not contribute in the $\epsilon$ regime. Similarly, let $Y^{(n)}_v$ 
and $Y^{(n)}_c$ be the value of the current susceptibilities when computed 
in the $n$ monomer sector. Expanding observables in the various monomer 
sectors it is possible to show
\begin{equation}
Y_i = Y_i^{(0)} + m^2\Bigg\{Y^{(2)}_i - Y^{(0)}_i\Bigg\} \frac{Z_2}{Z_0} + ...
\end{equation}
where $i=v,c$. From Eq.(\ref{chipred}) we see that
\begin{equation}
\frac{Z_2}{Z_0} = \Sigma^2\frac{L^{2d}}{8} \bigg(1 + \frac{3 \beta_1}{(FL)^2} +
\frac{a}{(FL)^4}  + ...\bigg)
\end{equation}
Since at $u=0$ no monomers contribute we also have
\begin{equation}
Y^{(0)}_i = \frac{F^2}{2} \Bigg(1 + \frac{\beta_1}{(FL)^2} +  
\frac{a'}{(FL)^4} + ...\Bigg)
\end{equation}
Substituting our knowledge of $Z_2/Z_0$ and $Y_c^{0}$ we get
\begin{equation}
Y_i = \Bigg(\frac{F^2}{2}\Bigg\{1 + \frac{\beta_1}{(FL)^2} +  ...\Bigg\} +
\frac{u^2}{8}\Bigg\{Y^{(2)}_i - Y^{(0)}_i\Bigg\} + ...\Bigg)
\end{equation}
Comparing with Eq.(\ref{ypred}) we conclude that
\begin{subequations}
\label{ypred1}
\begin{eqnarray}
Y^{(2)}_c &=& \frac{2 F^2}{3} \Bigg(1 +
\frac{3\beta_1}{2(FL)^2} +  \frac{b'_c}{(FL)^4} + ...\Bigg) \\
Y^{(2)}_v &=& \frac{F^2}{3} \Bigg(1 +  \frac{b'_v}{(FL)^4} + .. \Bigg)
\end{eqnarray}
\end{subequations}
Notice that $Y^{(2)}_c \neq Y^{(2)}_v$ due to the effects of explicit 
chiral symmetry breaking that is introduced due to the presence of the
monomers.

\begin{figure}[htb]
\vskip0.5in
\begin{center}
\includegraphics[width=12cm]{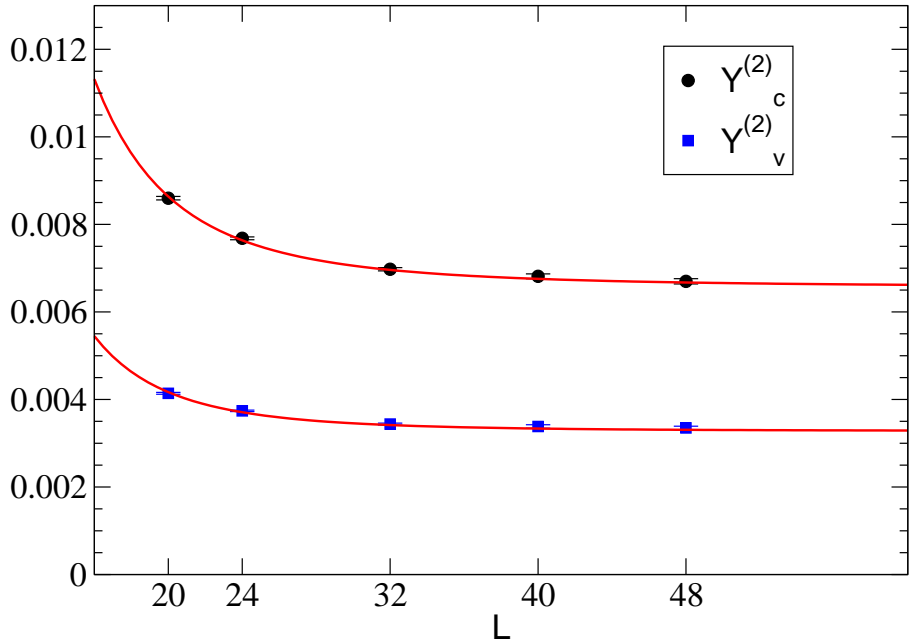}
\end{center}
\caption{\label{fig4} Plot of $Y^{(2)}_c$ and $Y^{(2)}_v$, evaluated
in the two monomer sector as a function of $L$ at $T=1.733$, $c=0.3$ 
and $m=0$. The solid lines are fits to Eq.~(\ref{ypred1}) discussed in
the text.}
\end{figure}

We have used the fixed monomer update to compute $Y_c^{(2)}$ and $Y_v^{(2)}$.
In Fig.(\ref{fig4}) we plot our results as a function of $L$. We fix 
$F=0.0992$ and try to fit our data to Eq.(\ref{ypred1}). We obtain
$b'_c=4.1(1)$ with a $\chi^2/DOF = 1.1$, and $b'_v= 4.2(1)$ with a
$\chi^2/DOF=2.1$. Although our results again appear consistent with the 
predictions at large $L$, the large values for the constants $b'_c$ 
and $b'_v$ show that we need data with small errors for $L>48$ to be 
sure we can be sensitive to the universal predictions at ${\cal O}(1/L^2)$.

While our model is analogous to real QCD, we can not expect our results 
to be equivalent to real QCD.  However, for pedagogical reasons, we can 
make comparisons to real QCD studies.  In particular, at $T=1.733$, we 
found $F a_{\rm lat} = 0.0992$ and $M_\eta a_{\rm lat} = 1.1$ where we
have introduced $a_{\rm lat}$ to be the lattice spacing. If we take 
$a_{\rm lat} = 1 GeV^{-1}$, we find $F \sim 99$MeV and $M_\eta \sim 1.1$GeV 
which are reasonably close to physical values.
%\cite{Fukaya:2007fb}.

\begin{figure}[htb]
\begin{center}
\includegraphics[width=12cm]{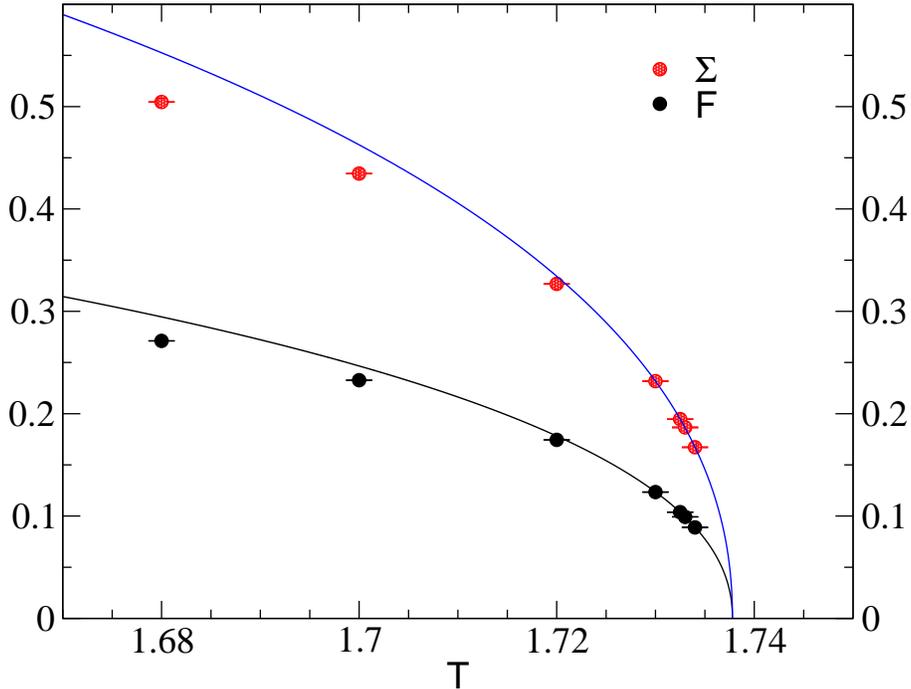}
\end{center}
\caption{\label{fig5} $F$ and $\Sigma$ plotted as a function of $T$. The
solid lines are fits to Eq.(\ref{eq:mfcrit}) with $A_F=0.943$,
$A_\Sigma=1.769$ and $T_c=1.73779$.}
\end{figure}

\section{Continuum Limit}
\label{climit}

At $c=0.3$ and $T=1.733$ we found that $F \sim 0.1$. Such a small value of
$F$ clearly is not an accident. At a fixed $c\neq 0$ the model should
go through a phase transition at some critical temperature $T_c$. If
this transition is second order then at the critical temperature one 
would expect $F$ to vanish. Thus, $T=1.733$ must be close to such a
critical point. Since we are in four dimensions, we expect quantities
to obey mean field scaling near a second order critical point. In particular
this means \cite{Zinn-Justin}:
\begin{subequations}
\begin{eqnarray}
F \sim A_{F}(T_c-T)^{\frac{1}{2}}|\log(T_c-T)|^{\frac{1}{4}} \\
\Sigma \sim A_{\Sigma}(T_c-T)^{\frac{1}{2}}|\log(T_c-T)|^{\frac{1}{4}}
\end{eqnarray}
\label{eq:mfcrit}
\end{subequations}
We have verified that our results for $c=0.3$ are consistent with this 
expectation. Tab.(\ref{Tabfits}) gives the chiral perturbation theory 
fit parameters as a function of $T$. The values of $F$ and $\Sigma$
are plotted as a function of $T$ in Fig.(\ref{fig5}). If we use the data 
for $T \geq 1.73$ a combined fit of $F$ and $\Sigma$ as a function of
$T$ to Eq.(\ref{eq:mfcrit}) gives $A_F=0.943(4)$, $A_\Sigma=1.769(4)$ 
and $T_c=1.73779(4)$ with a $\chi^2/DOF=0.7$.

\begin{table}
\begin{center}
\begin{tabular}{|c|| c c c|| c c c||c c |} \hline\hline
\em $T$ &\em $F$ &\em $a$ &\em $\chi^2$ & \em $\Sigma$ &\em $b$ &\em $\chi^2$ &\em $M_\eta$ &\em $\chi^2$
\\\hline
1.68  &0.2711(1) &1.28(3) &0.3 &0.5045(2) &1.22(4)  &1.1   &1.04(5) &1.5   \\ 
1.70  &0.2327(1) &1.91(9) &1.2 &0.4346(2) &1.72(11) &0.2   &1.06(5) &1.2   \\
1.72  &0.1744(1) &2.58(13)&0.1 &0.3268(2) &2.44(17) &0.5   &1.02(3) &1.1   \\
1.73  &0.1234(2) &2.25(10)&1.7 &0.2318(2) &2.26(8)  &1.5   &1.05(2) &1.6   \\
1.7325&0.1038(1) &2.80(18)&0.1 &0.1947(2) &2.77(23) &0.7   &1.08(3) &1.3   \\
1.733 &0.0992(1) &2.71(9) &0.8 &0.1866(4) &3.00(20) &1.8   &1.05(3) &1.5   \\
1.734 &0.0889(3) &2.78(15)&2.3 &0.1672(4) &2.81(17) &2.3   &1.08(3) &1.4   \\
\hline\hline
\end{tabular}
\caption{Fits of $Y_c$ and $\chi_\sigma$ as a function of $L$ based on 
Eqs.~(\ref{ypred}) and (\ref{chipred}) at $c=0.3$ and $m=0$ at various 
values of $T<T_c$. The last two columns give the value $M_\eta$ and 
the $\chi^2/DOF$ obtained from a fit to $G_\eta(x,y)$.\label{Tabfits}}
\end{center}
\end{table}

Although we cannot rule out a weak first order transition with a 
large correlation length, if we assume that there is a second order 
critical point at $T=T_c$, we can in principle make $F$ arbitrarily 
small by tuning our $T$ closer and closer to $T_c$. This raises an 
interesting question: By approaching $T_c$ can we define a 
``continuum limit'' of our lattice pion theory by holding $F$ 
to be the fixed physical scale? However, due to triviality of 
four dimensional scalar field theories this limiting theory is expected 
to be free \cite{Luscher:1987ek,Luscher:1988uq,Balog:2004zd}. On the other 
hand, chiral perturbation theory suggests that pions will always interact 
as their momentum increases which is in contradiction to the statement
of triviality. Hence, we conjecture that the mass of the $\sigma$ 
resonance will go to zero in units of $F$. Since triviality is a
logarithmic effect we expect $M_\sigma/F \sim 1/|\log(T_c-T)]|^p$ 
where $p$ is some power. This means that the chiral expansion would 
break down at momenta close to $M_\sigma$ rather than $F$ and the 
usual power counting of chiral perturbation theory would become 
questionable close to $T_c$. Thus, our approach to model pion physics 
of QCD should eventually fail! On the other hand since triviality is 
a logarithmic effect, it is likely that we still will have a window 
where our model produces pions very much like in two-flavor QCD. The 
present study is an attempt to demonstrate this in the $\epsilon$-regime.

\section{Conclusions and Future Work}

In this work we have developed a new approach to model the physics of
pions in $N_f=2$ QCD. Our approach uses $N_f=2$ lattice QED in the strong
coupling limit. We have shown that using the mapping to dimer models we 
can study our model very efficiently in the chiral limit and close to it. 
We have established consistency with the $\epsilon$ regime predictions of
chiral perturbation theory.  We have also demonstrated that we can make 
$F \ll 1$ by tuning a fictitious temperature so one approaches a 
second order phase transition. This tuning is expected to help remove 
lattice artifacts and approach a {\em continuum-like} theory. But, as 
explained in section \ref{climit}, our method will eventually fail due 
to logarithmic triviality of scalar field theories. Since triviality is 
a logarithmic effect, there should be a large window where our 
method may provide a model for pions. Our results confirm this belief.

There are many directions for the future. We are currently performing 
calculations in the $p$-regime and hope to understand the region in our
model where chiral perturbation theory, up to a particular order, will be 
valid \cite{Bijnens:1998fm,Colangelo:2002hy,Colangelo:2003hf,Colangelo:2004sc,
Colangelo:2004xr,Colangelo:2005cg,Colangelo:2005gd,Colangelo:2006mp}.
Since we have already computed the leading low energy constants in
the $\epsilon$-regime we can constrain them when we analyze the $p$-regime. 
Another interesting arena to explore is pion scattering and 
resonance physics by measuring the appropriate two and four point 
correlation functions and extracting scattering lengths and phase 
shifts via L\"{u}scher's method \cite{Luscher:1990ux,Beane:2005rj}. Our 
approach should also allow one to study the effects of the quark mass 
on these quantities.

\section*{Acknowledgments}

DJC gratefully achnowledges B.C. Tiburzi and F.J. Jiang for discussions
relevant to this work. SC thanks Gilberto Colangelo, Stephan D\"{u}rr, 
and Uwe-Jens Wiese for discussions and hospitality at Bern University 
where part of this work was done. DJC also acknowledges Robert G. Brown 
for technical assistance in C programming. This work was partially 
supported by the DOE grant DE-FG02-05ER41368.  

\bibliography{qcd}

\newpage

\section*{Appendix I}

Here we give the analytic expressions for the partition function, the 
chiral, vector, and axial helicity moduli and the two susceptibilities
for a $2\times 2$ lattice,as defined in Section(\ref{obs}).  (Note
$c = \tilde c + m^2$):
\begin{subequations}
\begin{eqnarray}
Z(T,c,m) &=& 36T^4 + 64T^2 + 36 + c^4 + 12(1+T^2)c^2 + 8(1+T)c^2m^2  
\\\nonumber
&+& 32Tcm^2 + 16(1+T^2)m^4 + 48(1+T^3)m^2 + 32(T+T^2)m^2 
\\
2Z \times Y^C_W(T,c,m) &=& 96 + 64T^2 + 16c^2 + 112m^2 + 32m^4 + 
8c^2m^2 + 32Tcm^2 \\\nonumber &+& 32(2T+T^2)m^2
\\
2Z \times Y^V_W(T,c,m) &=& 96 + 64T^2 + 16c^2 + 64m^2
\\
Y^A_W(T,c,m) &=& Y^C_W(T,c,m) + \frac{64T^2m^2}{2Z}
\\
2Z \times \chi_\pi &=& 24T^3 + 16T^2 + 16T + 24 + 4(1+T)c^2 + 
4(3+4T+3T^2)c + 2c^3 \\\nonumber &+& 8(1+T)cm^2 + 16(1+T+T^2)m^2\\
2Z \times \chi_\eta &=& 24T^3 + 16T^2 + 16T + 24 + 4(1+T+m^2)c^2 
- 4(3-4T+3T^2)c \\\nonumber &-& 2c^3 - 8(1+T)cm^2 + 8(5+6T+5T^2)m^2 
+ 16(1+T)m^4
\end{eqnarray}
\end{subequations}
In Tabs.(\ref{Tabhelic}) and (\ref{Tabsus}) we compare the analytic
results with the results obtained using our algorithm for different
parameters. The agreement gives us confidence that our algorithm must
be correct.

\begin{table}
\begin{center}
\begin{tabular}{c c c c c c c c c c} \hline\hline
&\em T &\em c &\em m & \em Algo. &\em Exact  & \em Algo. &\em Exact 
&\em Algo. &\em Exact\\\hline
& & & &$Y_w^A$ & &$Y_w^C$ & &$Y_w^V$ & \\
&1.0 &0.5 &0.0 &0.8023(9) &0.80246... &0.5763(6) &0.57721... 
&0.5771(8) &0.57721... \\ 
&1.5 &0.5 &0.0 &0.5212(7) &0.52141... &0.3275(6) &0.32790...
&0.3274(7) &0.32790...\\ 
&1.0 &1.0 &0.0 &0.7449(8) &0.74534... &0.5470(7) &0.54658...
&0.5468(8) &0.54658..\\ 
&1.5 &1.0 &0.0 &0.4967(7) &0.49720... &0.3178(5) &0.31821...
&0.3175(6) &0.31821...\\
&1.0 &1.5 &0.0 &0.6667(8) &0.66645... &0.5016(5) &0.50240...
&0.5014(6) &0.50240...\\
&1.5 &1.5 &0.0 &0.4607(6) &0.46147...&0.3029(5) &0.30325...
&0.3033(6) &0.30325...\\
&1.0 &2.0 &0.0 &0.5812(7) &0.58064... &0.4519(6) &0.45161...
&0.4521(6) &0.45161...\\
&1.5 &2.0 &0.0 &0.4199(5) &0.41927...&0.2853(5) &0.28451...
&0.2853(5) &0.28451...\\
&1.0 &0.0 &0.5 &0.7734(5) &0.77336... &0.5972(5) &0.59730...
&0.4867(6)  &0.48692... \\
&1.5 &0.0 &0.5 &0.5062(6) &0.50702... &0.3479(5) &0.34834...
&0.2827(4)  &0.28319... \\
&1.25 &0.5 &1.5 &0.4192(4) &0.41948... &0.3914(6) &0.39164...
&0.1285(3)  &0.12831... \\
&1.5 &1.25 &0.5 &0.4516(5) &0.45126... &0.3250(4) &0.32496...
&0.2564(4)  &0.25611... \\
&2.0 &1.25 &1.5 &0.2664(3) &0.26636... &0.2395(3) &0.23928... 
&0.0733(1) &0.07318...  \\
&1.0 &0.5 &2.0 &0.3834(5) &0.38411... &0.3757(4) &0.37622... 
&0.0915(2) &0.09122 
\\
\hline\hline
\end{tabular}
\caption{Helicity moduli for a $2\times2$ lattice as discussed in the text.\label{Tabhelic}}
\end{center}
\end{table}

\begin{table}
\begin{center}
\begin{tabular}{c c c c c c c c c } \hline\hline
& & & &$\chi_\pi$ & &$\chi_\eta$ & \\
&\em T &\em c &\em m & \em Algo. &\em Exact  &\em Algo. &\em Exact \\\hline
&1.0 &0.5 &0.0 &0.3601(3) &0.359877... &0.2172(2) &0.217334... \\ 
&1.5 &0.5 &0.0 &0.2676(2) &0.267764... &0.1824(1) &0.182429...\\ 
&1.0 &1.0 &0.0 &0.4040(3) &0.403727... &0.1429(2) &0.142857...\\ 
&1.5 &1.0 &0.0 &0.2981(2) &0.298322... &0.1367(1) &0.136731...\\
&1.0 &1.5 &0.0 &0.4220(3) &0.422301... &0.0801(1) &0.080103...\\
&1.5 &1.5 &0.0 &0.3172(3) &0.317264... &0.0948(1) &0.094767...\\
&1.0 &2.0 &0.0 &0.4194(2) &0.419355... &0.0323(1) &0.032258...\\
&1.5 &2.0 &0.0 &0.3251(3) &0.324754... &0.0588(1) &0.058961...\\
&1.0 &0.0 &0.5 &0.2849(2) &0.28481... &0.2849(2) &0.28481... \\
&1.5 &0.0 &0.5 &0.2220(1) &0.22220... &0.2220(1) &0.22220... \\
&1.25 &0.5 &1.5 &0.1708(2) &0.17081... &0.1412(2) &0.14121... \\
&1.5 &1.25 &0.5 &0.2767(2) &0.27650... &0.1175(1) &0.11753... \\
&2.0 &1.25 &1.5 &0.1384(1) &0.13831... &0.0933(1) &0.09323... \\
&1.0 & 0.5 &2.0 &0.1339(1) &0.13365... &0.1160(1) &0.11584... \\
\hline\hline
\end{tabular}
\caption{Susceptibilities for a $2\times2$ lattice as discussed in the text.\label{Tabsus}}
\end{center}
\end{table}

\end{document}